# Symmetric Algorithm Survey: A Comparative Analysis

Mansoor Ebrahim
IQRA University
Main Campus
Defense View, Karachi

Shujaat Khan
IQRA University
Main Campus
Defense View, Karachi

Umer Bin Khalid
IQRA University
Main Campus
Defense View, Karachi

## ABSTRACT
Information Security has become an important issue in modern world as the popularity and infiltration of internet commerce and communication technologies has emerged, making them a prospective medium to the security threats. To surmount these security threats modern data communications uses cryptography an effective, efficient and essential component for secure transmission of information by implementing security parameter counting Confidentiality, Authentication, accountability, and accuracy. To achieve data security different cryptographic algorithms (Symmetric & Asymmetric) are used that jumbles data in to scribbled format that can only be reversed by the user that have to desire key.

This paper presents a comprehensive comparative analysis of different existing cryptographic algorithms (symmetric) based on their Architecture, Scalability, Flexibility, Reliability, Security and Limitation that are essential for secure communication (Wired or Wireless).

## 1. INTRODUCTION
Cryptography a Modern encryption technology, comprising of different mathematical processes involving the application of formulas (or algorithms) was conventionally designed to secure discretion of military and diplomatic communications. With the Rapid growth of information technology and science of encryption, an innovative area for cryptographic products has stimulated. Cryptography [1] is defined as "the subdivision of cryptology in which encryption /decryption algorithms are designed, to guarantee the security and authentication of data". Cryptography can be classified as Symmetric key algorithm and Asymmetric key algorithm.

Symmetric-key algorithms [2] also known as single- key, one-key and private-key encryption are a class of algorithms for cryptography, that uses a Private(shared secret) key and a Public (non-secret) algorithm to execute encryption /decryption process. Some popular and well-respected symmetric algorithms includes DES [3], TDES [4], Blowfish [5], CAST5 [6], IDEA [7], TEA [34], AES (aka Rijndael) [8, 9, 10, 11, 12, 13], Twofish [8] [14], RC6, Serpent and MARS. Asymmetric-key algorithms [2] also known as public key encryption is a form of crypto system in which encryption and decryption are modern encryption technology mathematically performed using different keys, one of which is referred to as public key and the other is referred to as private key. Some popular and well-respected asymmetric algorithms includes PGP [4] (Pretty Good Privacy, with versions using RSA [15] and Diffie-Hellman keys [16], SSH [4] (the secure alternative to telnet) and SSL [4] (used for encryption of data between a web browser and server). This paper provides an overview, detail evaluation and analyses of existing symmetric cryptographic algorithms.

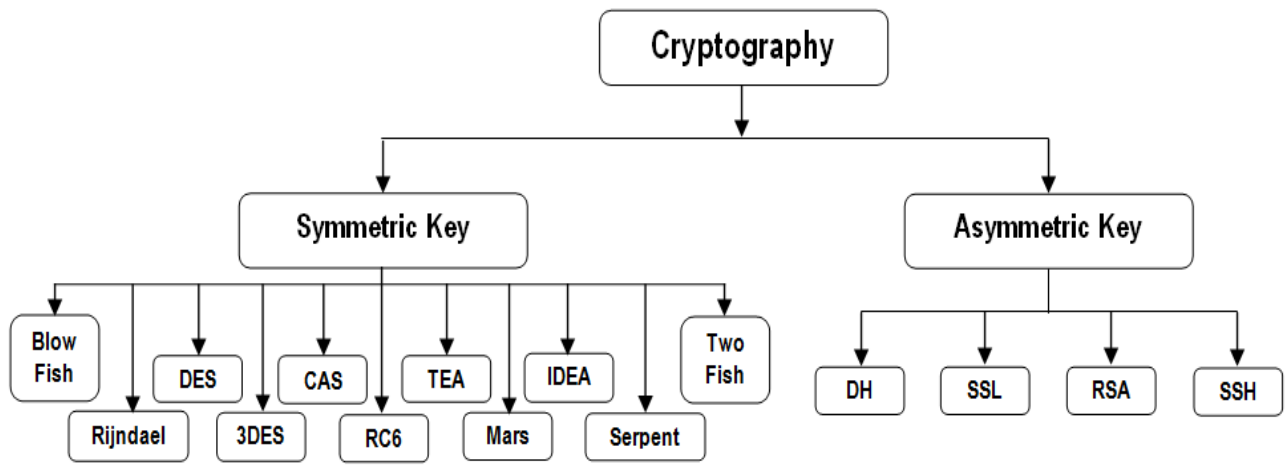

Figure 1: Different Symmetric & Asymmetric Cryptographic Algorithms

## 2. SYMMETRIC ALGORITHMS

In this section, different types of existing symmetric algorithms have been evaluated. In order to apply an appropriate algorithm in a particular application it is required to know its strength and limitation. Consequently the assessment of different existing algorithms based on certain parameters is necessary. The parameters may include Architecture, Security, Scalability (in terms of Encryption rate, Memory Usage, Software hardware performance and computational time), Limitations, and Flexibility.

### 2.1 Criteria

The criteria on which the different algorithms are being analyzed are;

#### *2.1.1 Architecture*
Defines the structure and operations that an algorithm can perform, its characteristics and how they are implemented. It also determines that the algorithm is symmetric or asymmetric that is whether it makes use of secret key or public key for encryption and decryption.

#### *2.1.2 Security*
An affirmative measure of the system strength in resisting an attack is a desirable element of any encryption algorithm possesses the property of in distinguishability (built by combining substitution with transposition repeatedly). Security of an encryption algorithm depends on the key size used to execute the encryption: generally, greater the keys size stronger the encryption. Length of key is measured in bits.

#### *2.1.3 Flexibility*
Defines whether the algorithm is able to endure minor modifications according to the requirements.

#### *2.1.4 Scalability*
It is one of the major element on which encryption algorithms can be analyzed. Scalability depends on certain parameters such as Memory Usage, Encryption rate, Software hardware performance; Computational efficiency.

#### *2.1.5 Limitations (Known Attacks)*
Defines how fine the algorithm works by make use of the computer resources available to it. Further how often is vulnerable to different types of attacks.

### 2.2 Assessment Methodology

The evaluation methodology was easy and simple, different encryption algorithm codes were downloaded as well as resources such as manuals, source code and research papers were studied, each algorithm was evaluated on the basis of above-mentioned parameters. Not a single algorithm fully contented the evaluation criteria, with some having greater deficiencies than others.

Authors of different algorithms claims to the scalability of their algorithms, simulations were carried out on the scalability of different algorithms [1-3].

## 3. ALGORITHM REVIEW

### 3.1 Overview

#### *3.1.1 DES*
Data Encryption Standard (1974), designed by IBM [17] based on their Lucifer cipher was the first encryption standard to be published by NIST (National Institute of Standards and Technology).The DES was initially considered as a strong algorithm, but today the large amount of data and short key length of DES limits its use [3].

#### *3.1.2 Triple-DES*
DES was superseded by triple DES (3DES) in November 1998, concentrating on the noticeable imperfections in DES without changing the original structure of DES algorithm.

TDES was much more complicated version of DES achieving high level of security by encrypting the data using DES three times using with three different unrelated keys.3DES [18] is still approved for use by US governmental systems, but has been replaced by the advanced encryption standard (AES) Sub subsections [4].

#### *3.1.3 Blowfish*
Blowfish by Bruce Schneier, author of Applied Cryptography, is considered as a highly rated encryption algorithm in terms of security, with different structure and Functionality than the other mentioned encryption algorithms.

Blowfish is a fast, compact, and simple block encryption algorithm with variable length key allowing a tradeoff between speed and security. Blowfish is a public domain algorithm (unpatented) and is used in the SSL and other program [5].

#### *3.1.4 IDEA*
James L. Massey and Xuejia Lai (Zurich, Switzerland) in 1990 developed an encryption algorithm named as International Data Encryption Algorithm (IDEA). It is fairly fast, considered secure, and is also resistant to both linear and differential analysis. IDEA [19] is considered one of the secure block ciphers offered in public domain in last decades [7].

#### *3.1.5 TEA*
David Wheeler and Roger Needham (Cambridge Computer Laboratory) in 1994 designed TEA, first presented and published in the proceedings at the Fast Software Encryption workshop. The Tiny Encryption Algorithm (TEA) is known for its simple structure and easy implementation, typically a few lines of code [34].

#### *3.1.6 CAST 5*
CAST 5 (1996) was produced by Carlisle Adams and Stafford Tavares. In cryptography, CAST-128 (CAST 5) is a block cipher used for different applications, particularly as an evasion cipher in various editions of GPG and PGP. It is also used by the Canadian Communications Security Establishment permitted by Canadian government [6].

#### *3.1.7 AES (Rijndael)*
Rijndael developed by Joan Daemen and Vincent Rijmen, becomes U.S.'s new Advanced Encryption Standard in October 2000 declared by the National Institute of Standards and Technology. Rijndael using variable key size is extremely fast and compact cipher. Its symmetric and parallel structure provides great flexibility for implementers, with effective resistance against cryptanalytic attacks [18].

AES can be well adapted to a wide range of modern processors such as Pentium, RISC and parallel processors. In general, AES is the name of the standard, and Rijndael is the algorithm described however, in practice the algorithm is also referred to as "AES".

### 3.1.8 AES (RC6)

RC6 a derivative of RC5, designed by Ron Rivest, Matt Robshaw, Ray Sidney, and Yiqun Lisa Yin [20] is a symmetric key algorithm. It was design to congregate the requirements of the Advanced Encryption Standard (AES) contest and was selected as one of the five finalists, and was also presented to the NESSIE and CRYPTREC projects.

It is patented by RSA Security [21]. RC6 offers good performance in terms of security and compatibility.

### 3.1.9 AES (Serpent)

Serpent another finalist of Advanced Encryption Standard (AES) [20] competition, stood 2nd to Rijndael, is a symmetric key block cipher, designed by Ross Anderson, Eli Biham, and Lars Knudsen. Security presented by Serpent was based on more conventional approaches than the other AES finalists. The Serpent is open in the public sphere and not yet patented.

### 3.1.10 AES (Two Fish)

Bruce Schneier along with John Kelsey, Doug Whiting, David Wagner, Chris Hall, and Niels Ferguson; extends the Blowfish team to enhance the earlier block cipher Blowfish to its modified version named Twofish to met the standards of AES for algorithm designing.

It was one of the five finalists of the Advanced Encryption Standard contest, but was not selected for standardization. The Twofish [22] is an open to public sphere and not yet patented.

### 3.1.11 AES (MARS)

The MARS base on layered, compartmentalized approach included Don Coppersmith (DES team member). MARS was exclusively designed to resist future advances in cryptography. MARS is a block cipher submitted by IBM's Advanced Encryption Standard contest and was selected as one of the five finalists and was given the last position among the five finalists, the permutation of elevated security, speed, and flexibility, makes MARS an exceptional alternative for the encryption needs of the information world.

## 3.2 Architecture

### 3.2.1 DES

DES is symmetric key algorithm based on the backbone concept of Feistel Structure. The DES is a block cipher that uses a 64 bit plain text with 16 rounds and a Key Length of 56-bit, originally the key is of 64 bits (same as the block size), but in every byte 1 bit in has been selected as a 'parity' bit, and is not used for encryption mechanism.

The 56 bit is permuted into 16 sub- keys each of 48- bit length. It also contains 8 S- boxes and same algorithm is used in reversed for decryption [3].

### 3.2.2 Triple DES

3DES is exactly what it is named–it performs 3 iterations of DES encryption on each block. As it is an enhanced version of DES so is based on the concept of Feistel Structure. The 3DES uses a 64 bit plain text with 48 rounds and a Key Length of 168-bits permuted into 16 sub- keys each of 48- bit length. It also contains 8 S- boxes and same algorithm is used in reversed for decryption [4].

### 3.2.3 Blowfish

Blowfish is also a symmetric key Feistel Structured algorithm consisting of 2 parts: key expansion part and data-encryption part. Blowfish is a block cipher that uses a 64 bit plain text with 16 rounds, allowing a variable key length, up to 448 bits, permuted into 18 sub- keys each of 32- bit length and can be implemented on 32- or 64-bit processors. It also contains 4 S-boxes and same algorithm is used in reversed for decryption [5].

### 3.2.4 IDEA

IDEA is symmetric key algorithm based on the concept of Substitution-Permutation Structure. It is a block cipher that uses a 64 bit plain text with 8 rounds and a Key Length of 128-bit permuted into 52 sub- keys each of 128- bits. It does not contain S- boxes and same algorithm is used in reversed for decryption [7].

### 3.2.5 TEA

TEA is also a Feistel Structured symmetric key algorithm. TEA is a block cipher that uses a 64 bit plain text with 64 rounds and a Key Length of 128-bit with variable (recommended 64 Feistel rounds) rounds having 32 cycles. It does not contain S- boxes and same algorithm is used in reversed for decryption [34].

### 3.2.6 CAST

CAST is symmetric key algorithm based on the backbone concept of Feistel Structure. The CAST is a block cipher that uses a 64 bit plain text with 12 or 16 rounds and a variable Key Length of 40 to128-bit. It also contains 4 S- boxes and same algorithm is used in reversed for decryption [6].

### 3.2.7 AES (Rijndael)

AES is also a symmetric key algorithm based on the Feistel Structure. The AES is a block cipher that uses a 128 bit plain text with variable 10, 12, or 14 rounds (Rijndael's Default # of Rounds is dependent on key size. Default # of Rounds = key length/32 + 6) and a variable Key Length of 128, 192, 256 bit permuted into 10 sub- keys each of 128, 192, 256 bit length respectively. It only contains a single S- box and same algorithm is used in reversed for decryption.

### 3.2.8 AES (RC6)

RC6 is a Feistel Structured private key algorithm that makes use a 128 bit plain text with 20 rounds and a variable Key Length of 128, 192, and 256 bit. As RC6 works on the principle of RC, can sustain an extensive range of word-lengths, key sizes and number of rounds, RC6 [21] does not contain S- boxes and same algorithm is used in reversed for decryption.

### 3.2.9 AES (Serpent)

Serpent is a symmetric key algorithm that is based on substitution-permutation network Structure. It consists of a 128 bit plain text with 32 rounds and a variable Key Length of 128, 192, and 256 bit. It also contains 8 S- boxes and same algorithm is used in reversed for decryption.

### 3.2.10 AES (Twofish)

Twofish is also a symmetric key algorithm based on the Feistel Structure. The AES is a block cipher that uses a 128 bit plain text with 16 rounds and a variable Key Length of 128, 192, 256 bit. It makes use of 4 S-boxes (depending on Key) and same algorithm is used in reversed for decryption.

### 3.2.11 AES (MARS)

MARS is symmetric key algorithm based on heterogeneous Structure. MARS make use of a 128 bit plain text with 32 rounds and a variable key length from 128 to 448 bits (multiple of 32-bit). It only contains a single S- box the same algorithm is used in reversed for decryption.

Table 1. Summary of Symmetric Algorithms Architecture

|  | Algorithm Structure | Plain Text/Cipher Text Length | Key Size | # S boxes | # of Rounds |
|---|---|---|---|---|---|
| **DES** | Festial structure | 64 bits | 56 | 8 | 16 |
| **3DES** | Festial structure | 64 bits | 168 | 8 | 48 |
| **Blowfish** | Festial structure | 64 bits | 128-448 | 4 | 16 |
| **IDEA** | Substitution-Permutation Structure | 64 bits | 128 | N/A | 8 |
| **TEA** | Festial structure | 64 bits | 128 | N/A | 64 (32 cycles) |
| **CAST** | Festial structure | 64 bits | 40-128 | 4 | 12 – 16 |
| **Rijndael** | Festial structure | 128 Bits | 128,192,256 | 1 | 10,12,14 |
| **RC6** | Festial structure | 128 Bits | 128,192,256 | N/A | 20 |
| **Serpent** | Festial structure | 128 Bits | 128,192,,256 | 8 | 32 |
| **Twofish** | Festial | 128 Bits | 128,192,256 | 4 | 16 |
| **MARS** | Festial | 128 Bits | 128-448 | 1 | 32 |

## 3.3 Security

### 3.3.1 DES

The security strength of DES depend on its 56 bit key size generating 7.2 x 1016 possible keys, making it extremely difficult to originate a particular key in typical threat environments. Moreover, if the key is changed frequently, the risk of unauthorized computation or acquisition can be greatly moderated. Moreover DES exhibits a strong avalanche effect i.e. a miniature modification in the plaintext or key, might change the cipher text noticeably. Initially DES was considered secure and was difficult to crack; Brute-force attacks became a subject of speculation immediately after the release of algorithm's in public domain, although DES survives different linear and differential attacks but in 1998 Electronic Frontier Foundation (EFF) designed a special-purpose machine for "decrypting DES". In one demonstration, it achieves the key of an encrypted message [23] in less than a day in combination with an alliance of computer users all around the world.

In general DES was proved insecure for large corporations or governments and it is simpler not to use DES algorithm. However for backward compatibility, and cost of upgrading, DES should still be preferred, outweighing the risk of exposure

### 3.3.2 TDES

TDES is an enhanced version of DES; 3DES use a larger size of key (i.e. 168-bits) to encrypt than that of DES. DES operations (encrypt-decrypt-encrypt) are performed 3 times in 3DES with 2-3 different keys, offering "112 bits of security" , avoiding so-called meet-in-the-middle attack [24].

TDES offers high level of security in comparison with DES and still in use by the US government.

### 3.3.3 Blowfish

Blowfish's security lies in its variable key size (128-448 bits) providing high level of security, Attempts to cryptanalysis Blowfish started soon after its publication however less cryptanalysis attempts were made on Blowfish than other algorithms. Blowfish is invulnerable against differential related-key attacks, since every bit of the master key involves many round keys that are very much independent, making such attacks very complicated or infeasible. Such autonomy is highly enviable.

### 3.3.4 IDEA

IDEA has a strong resistance against differential cryptanalysis under certain hypothesis. IDEA makes use of multiple group operations [25] to increase its strength against most familiar attacks. IDEA [26] consists of 128 bit key size making it a strong security algorithm. No weaknesses relating linear or algebraic attacks have yet been reported. The best attack which applies to all keys can break IDEA reduced to 6 rounds.

### 3.3.5 TEA

TEA algorithm offers the same security level as that of IDEA, it also consist of a 128 bit key size and is known for its simple structure and easy implementation.

### 3.3.6 CAST

CAST make use of variable key size operation to increase its security strength, the security of CAST is of great level and it resistant against both linear & differential attacks.

### 3.3.7 AES (Rijndael)

Security of Rijndael depends on its variable nature key size allowing up to a key size of 256-bit, to provide resistance against certain future attacks (collision attacks and potential quantum computing algorithms) [24].

General attacks that were revealed against concentrated rounds editions of Rijndael [25] are Square Attack, Improved Square Attack, Impossible Differential Attack and Reversed Key Schedule Attack, but none of the attacks were practically possible.

### 3.3.8 AES (RC6)

RC6 security lies in the completely random series of its output bits with 15 rounds or less [25] , running on input blocks of 128 bits, one of the parameter to make an encryption algorithm resistant against the attacks is that its output follows an entirely random series of bits [25]. A linear cryptanalysis attack can be launched for 16 rounds RC6, but requires 2^119 known plaintexts, which make the feasibility of such attack impossible. The RC6 algorithm is also strong against differential cryptanalysis, which worked with more than 12 rounds.

### 3.3.9 AES (Serpent)

Serpent is based on more conventional security approaches than the other AES finalists, opting a larger security margin. According to the author of Serpent 16 rounds Serpent quite adequate against all known types of attack [27], but as an indemnity against future discoveries in cryptanalysis it is extended to 32 rounds. In order to avoid the collision attack [27] Serpent usually discreet to modify keys well before 264 blocks have been encrypted [27]. Serpent with its minimum potential (only half number of rounds) is still as secure as that of three-key triple DES [27].

### 3.3.10 AES (Twofish)

Twofish algorithm is considered as robust and highly resistive to related key attacks including slide attack and the related key differential attack [25], with no weak keys that can be used to launch any related key attack.

### 3.3.11 AES (MARS)

MARS offers enhanced security and speed than triple DES and DES. It is an iterated cipher with unusually 32 rounds of different types. The middle rounds of MARS are the considered as its strong part. The security of MARS is dependent on data rotations (or functions with Boolean complexity). So, Visual Cryptanalysis is not successful against MARS. MARS algorithm is highly resistant to against all kind of Relative key attacks, Differential attacks and timing attacks.

## 3.4 Scalability

In this section Scalability of different algorithms are analyzed on the basis of memory usage and encryption performance (encryption and key scheduling).

The memory usage can be defined as the number of functions performed by the algorithm, smaller the memory usage greater will be the efficiency. Encryption rate is the processing time required by the algorithm for certain data size. Encryption rate is dependent on the processor speed, and algorithm complexity etc. The smallest value of encryption rate is desired. The hardware and software must be in accordance with the algorithm for better performance.

The graph in figure 2a and 2b shows generic scalability (memory usage & encryption performance) of different algorithm, the analysis are derived from different research papers. The paper [30] by Bruce Schenier provides a comprehensive analysis of the performance of the five AES finalist showing approximated algorithm speed against on a variety of common software and hardware platforms. In papers [31, 32 and 33] algorithm mentioned in Figure 2a are analyzed on the basis of memory usage, encryption performance and hardware implementation issues on common platforms, summarizing the overall scalability performance of 11 popular symmetric algorithms extracted from different research papers.

The graph in figure 2a shows the comparison between all the algorithms that were designed before AES whereas graph in figure 2b comprises of five algorithm that where AES finalists. If we compare both the graphs provided in figure 2a &b it can be observed that TEA algorithm is best among all the other existing algorithm in terms of encryption performance (High) and memory usage (Minimum). But its security has been compromised [32] so it is currently obsolete.

So, it is concluded that for most operational systems scalability is simply another parameter that must be incorporated in to a design and must be trade off with other features (Security, architecture, flexibility and robustness). It is very difficult to compare cipher designs for scalability and even more difficult to design cipher that are scalable among all platforms.

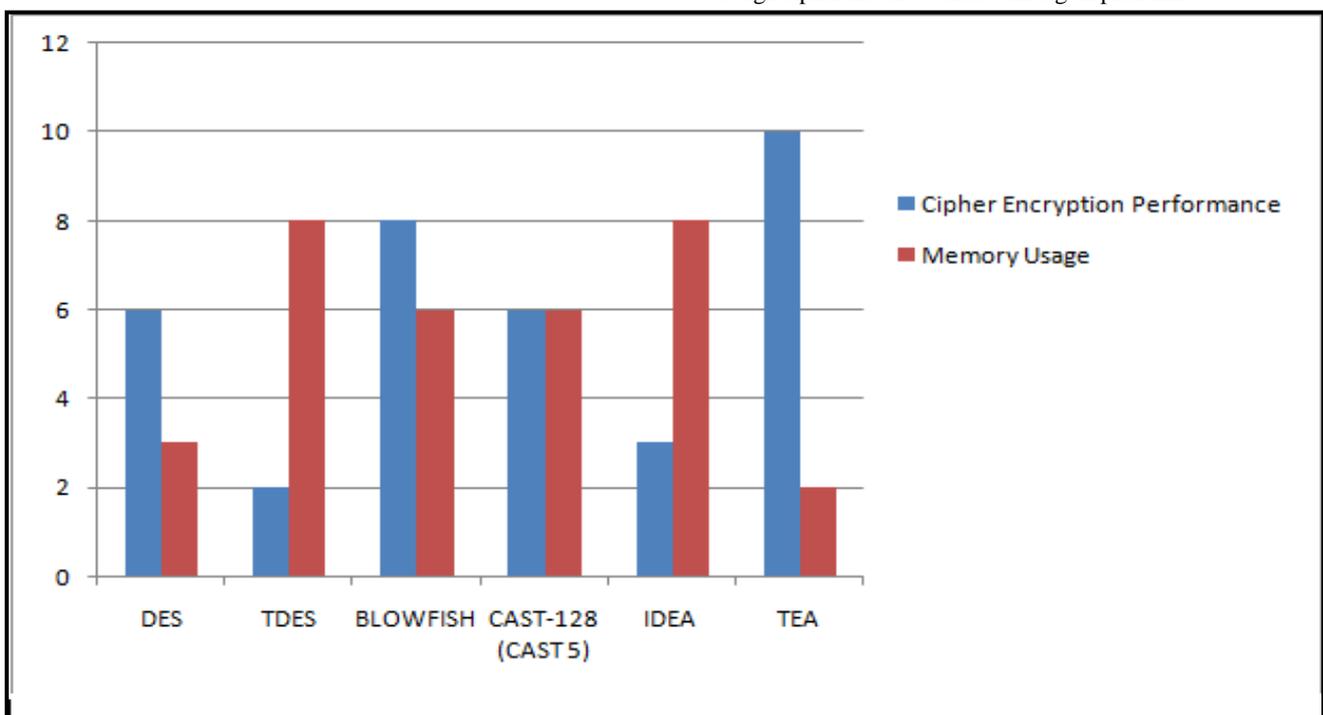

**Fig 2a: Generic Scalability (Memory Usage & Encryption Performance) of Different Algorithm**

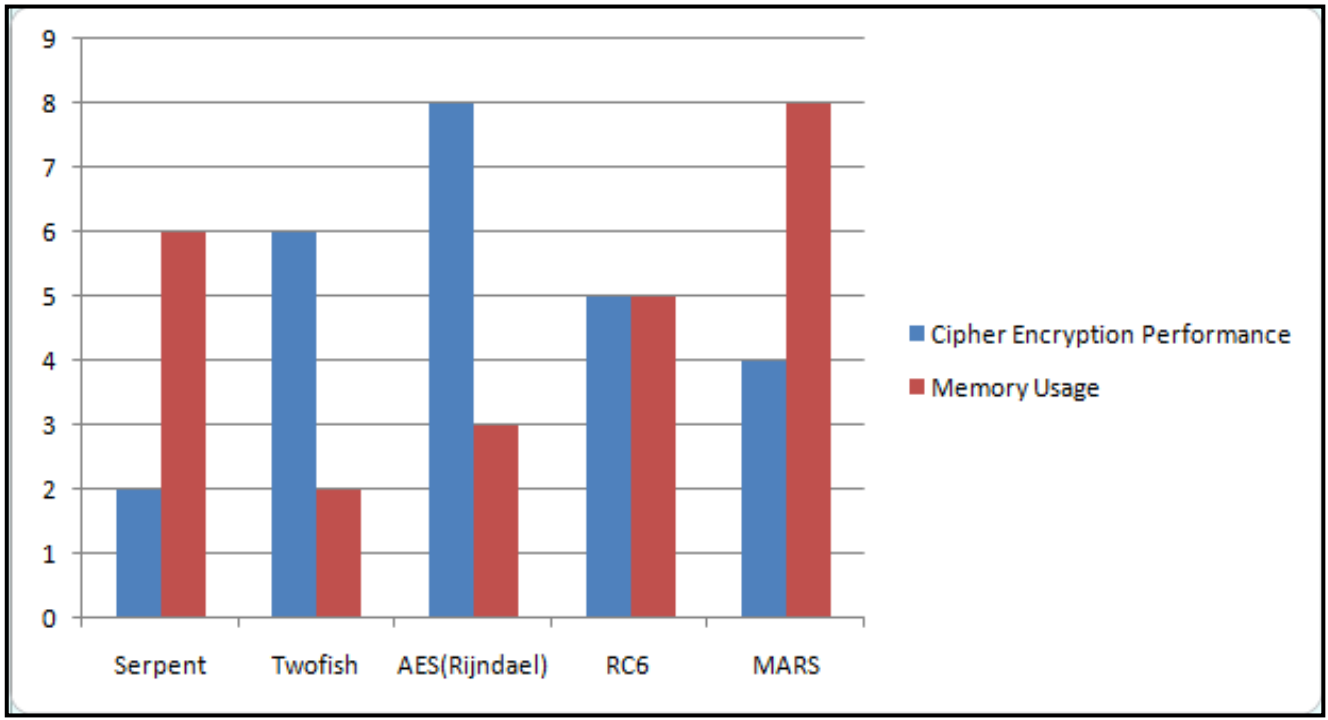

**Fig 2b: Generic Scalability (Memory Usage & Encryption Performance) of Different Algorithm**

## 3.5 Flexibility

In table 2 different algorithm are analyzed on the bases of their flexibility i.e. the ability of an algorithm to accept modifications according to the requirements

**Table 2. Summary of Symmetric Algorithms Flexibility**

| Algorithms | Flexible | Modification | Comments |
|---|---|---|---|
| **DES** | No | none | The structure of DES doesn't support any modifications |
| **3DES** | Yes | 168 | The structure of 3DES is same as DES ,it doesn't support any changes but as it iterates DES 3 times so the key size is extended to 168 bits |
| **Blowfish** | Yes | 64-448 | Blowfish key length must be multiples of 32 bits |
| **IDEA** | No | none | The structure of IDEA doesn't support any modifications |
| **TEA** | No | none | The structure of TEA doesn't support any modifications |
| **CAST** | Yes | 64,128,256 | 64 bit Cast was too expose to different type of linear & differential attacks, due to its flexible structure it was modified to 128 and 256 bits, increasing its security and strength. |
| **Rijndael** | Yes | 128,192,256 | The structure of AES(R) was extendable to the multiple of 64 bits, have same sub key size as the size of the key |
| **RC6** | Yes | 128-2048 | RC6 has a variable key length and can be extended to 2048 bits however the key lengths must be a multiple of 32 bits |
| **Serpent** | Yes | 256 | Serpent keys are always padded to 256 bits. The padding consists of a "1" bit followed by "0" bits. |
| **Twofish** | Yes | 256 | Two fish keys, other than the default sizes, are always padded with "0" bits up to the next default |
| **MARS** | Yes | 128-448 | MARS operates with variable key lengths, but the key length must be multiples of 32 bits |

### 3.6 Limitation

*3.6.1 DES:*
DES is highly vulnerable to linear cryptanalysis attacks, Weak keys is also a great issue. DES is also exposed to brute force attack [25].

*3.6.2 3DES:*
3 DES is exposed to differential and related-key attacks. It is also susceptible to certain variation of meet-in-the-middle attack [25].

*3.6.3 Blowfish*
Blowfish has some classes of weak keys. 4 rounds of blowfish are exposed to 2nd order differential attacks. So, reliability of Blowfish is questionable due to the large no. of weak keys [25].

*3.6.4 IDEA*
Some susceptibility regarding different classes of weak keys and minimum rounds version were observed in IDEA. It is also exposed to collision attack. IDEA contains 8 rounds in which first 3 rounds appears to highly exposed to key attacks such as key-schedule attacks and related-key differential timing attacks [25].

*3.6.5 TEA*
The major problem with TEA algorithm is Equivalent keys in which each key is equivalent to three others reducing the effective key size to a minimum of 126 bits. Further it is also exposed to related key attack involving $2^{23}$ chosen plain texts under a related-key pair, with complexity of $2^{32}$ [25].

*3.6.6 CAST 5*
The reduce version of CAST (64-bit key version) is susceptible to differential related-key attack [25]. It can be broken by $2^{17}$ chosen plaintexts along with one related-key query in offline work of $2^{48}$ [25].

*3.6.7 AES (Rijndael)*
AES(R) has no serious weakness; although it was observed that a mathematical property (not an attack) of the cipher might be vulnerable into an attack. Further in AES (Rijndael) the inverse cipher implementation is inappropriate on a smart card than the cipher itself [25].

*3.6.8 AES (RC6)*
In RC6, for a single class of weak keys, it is observed that full arbitrariness is not achieved for up to 17 rounds of the algorithm [25]. No other limitations were identified [25].

*3.6.9 AES (Serpent)*
No such limitation was found in serpent; however the 32 rounds make Serpent a bit slower and complex to implement on small blocks[25].

*3.6.10 AES (Twofish)*
Twofish is possibly susceptible to chosen-key attacks that may reduce the security of algorithm when applied to certain implementations, such as a hash function [25].

*3.6.11 AES (MARS)*
No significant limitations in MARS were observed. Due to different component natures involved in MARS the simple round function of MARS becomes relatively complex to analyze. The implementation of MARS on hardware is a bit difficult and complex [25].

### 4. COMPARATIVE ANALYSIS
After analyzing the most popular symmetric algorithms AES(Rijndael) was found the most secure, faster and better among all the existing algorithm with no serious weaknesses, there are some flaws in symmetric algorithms such as weak keys, insecure transmission of secret key, speed, flexibility, authentication and reliability i.e. in DES, four keys for which encryption is exactly the same as decryption [17]. This means that Original plain text can be recovered, if the encryption is applied twice with one of these weak keys [17]. DES is very slow when implemented in software; the algorithm is best suited to implementation in hardware. Similar is the case in IDEA that involves large class of weak keys facilitating the cryptanalysis for recovering the key. DES and IDEA have the same encryption speed on. Triple DES does not always provide the extra security that might be expected making use of double and triple encryption as well as it is very slow when implemented in software as it is derived from DES and DES on software is already slow, so Triple-DES might be considered safest but slowest. In Blow Fish there are certain weak key that attacks its three-round version, further it is also exposed to a differential attack against its certain variants, it also slow in speed but much more faster than DES and IDEA. While looking at the five finalists of AES no serious weakness was found, however few feeble aspect was highlighted that might be exploit as a molest in near future, such as in AES(Rijndael) a numerical property of the cipher might be exposed into an attack, full RC6 arbitrariness is not achieved, Serpent a bit slower and complex, Twofish possibly suspected to chosen-key attacks and MARS relatively complex to analyze.[28] [29].

### 5. CONCLUSION
In this paper a detailed analysis of symmetric block encryption algorithms is presented on the basis of different parameters. The main objective was to analyze the performance of the most popular symmetric key algorithms in terms of Authentication, Flexibility, Reliability, Robustness, Scalability, Security, and to highlight the major weakness of the mentioned algorithms, making each algorithm's strength and limitation transparent for application. During this analysis it was observed that AES (Rijndael) was the best among all in terms of Security, Flexibility, Memory usage, and Encryption performance. Although the other algorithms were also competent but most of them have a tradeoff between memory usage and encryption performance with few algorithms been compromised.

### 6. ACKNOWLEDGMENTS
The authors would like to acknowledge the support of Sir Syed University of Engineering & Technology and IQRA University Karachi.